# Static structures of strained carbon chains: DFT-modeling vs classical modeling of the chain with Lennard-Jones potential

G.M. Chechin[†] and V.S. Lapina
[†]e-mail: gchechin@gmail.com

Southern Federal University, Institute of Physics, Stachki Ave., 194, 344090, Rostov-on-Don, Russia

We proved earlier that in the strained monoatomic chains with Lennard-Jones potential there can exist an equilibrium static bi-structure, which corresponds to $N - 1$ equal short interatomic bonds and one long bond with inversion in its center ($N$ is the number of atoms of the chain). In the present work, we investigate with the aid of the density functional theory (DFT modeling) similar structures that can exist in the strained carbon chains. In contrast to the Lennard-Jones model, the bi-structures in this case are *inhomogeneous* (they have short bonds of different lengths) and appear abruptly when the strain exceeds a certain critical value $\eta_c$ as a result of a hard bifurcation of the equilibrium state (an analogue of the first-order phase transition). Such a bifurcation is associated with two different parts of potential, in which the Lennard-Jones forces can be quite small (this is the vicinity of the potential minimum and the potential tail) and, therefore, the forces acting on the particle between the long and short bonds can be equal. Since practically all interatomic potentials possess such features, the above bifurcation is universal, i.e. it must occur for any monoatomic chain. We have studied in detail the properties of the above structures, in particular, the behavior of their parameters with increasing $N$. With the help of DFT-modeling, the electron density of the above structures near the bifurcation point is investigated.

## 1. Introduction

According to theoretical predictions, individual monoatomic carbon chains, carbynes, should have many unique physical properties, which represent great interest for fundamental and applied science. Carbynes can exist in two different forms – cumulene with equal interatomic bonds and polyyne with alternation of short and long bonds. At the present time, the chemical synthesis of carbynes and their experimental study encounters great technical difficulties and, therefore, theoretical methods of their studying are very important. Most of these methods are based on the density functional theory [1-3], implemented in a number of powerful computational packages, such as ABINIT, Quantum Espresso, VASP, and others. Many interesting results were obtained in this way [4-11].

In [6], for long strained carbyne chains with even number of atoms, the Peierls phase transition was predicted above a certain threshold of the strain. As a result of this transition, carbyne transforms from metallic state to insulator state, and one can use this property in nanodevices to control the conductivity of the material by mechanical action. In [11], large amplitude nonlinear oscillations were studied and a sharp softening of the pi-mode frequency was found above a certain critical value of the strain. Condensation of this mode also leads to the Peierls transition. Moreover, the concept of soft modes allowed the authors to suggest, that there can exist two new forms of carbyne in nature, which differ from the polyyne in the type of alternation of short and long chemical bonds. In the same paper, a simple classical model was proposed which allowed qualitatively, and even quantitatively, in a number of calculations, to explain the above softening of nonlinear normal modes for appropriate values of the strain. This model represents a monoatomic chain whose interparticle interactions are described by the Lennard-Jones potential. Hereafter we refer to it as the L-J model.

Properties of sufficiently short carbon chains are discussed in [4, 5, 9]. To passivate chemically active ends of these chains a number of hydrogen atoms can be attached to their ends. If two hydrogen atoms are attached to each end of the chain, then the bond lengths in its middle part correspond approximately to the cumulene structure. If at each end of the chain there is only one hydrogen atom, then the polyyne structure appears. The method for identifying various forms of carbyne by infrared absorption spectra is discussed in [9]. Here, it is appropriate to note that carbyne was detected in galaxies dusty clouds by precisely optical methods [12].

The static structures of the *short* strained carbon chains were studied in detail in [4, 5].

In particular, these works discuss the fact that distribution of bond lengths and many other properties of such chains depend on the *parity* of the number of atoms in the chain. For example, chains with an *odd* number of atoms possess greater tensile strength.

Particular interest represents the studying of *oriented* carbyne, which is an ensemble of short carbon chains perpendicular to the substrate surface to which they are attached with their hydrogen ends [13, 14, 15]. It is argued in [15] that this material can be a topological insulator with two-dimensional superconductivity.

In [16], we have predicted for strained carbon chains the possibility of existence of a certain static *bi-structures*, in whose vicinity discrete breathers of a new type can exist. This study was done for the Lennard-Jones model. In the present work, we investigate the possibility of existence of similar structures in the strained carbon chains in the framework of the density functional theory, i.e. with the aid of DFT-modeling.

## 2. Some calculation details

All calculations in the present work were carried out within the framework of the density functional theory [1-3] with the application of the well-known computational packages [17, 18] (ABINIT 8.0.8, Quantum Espresso 6.1, etc.), which implemented the computational methods of this theory, and calculations for the Lennard-Jones model were performed with the aid of the mathematical package Maple 11. We used some conventional approaches such as the local density approximation (LDA), pseudo-potentials by Troullier-Martins to describe the field of the carbon atoms inner shells. Plane waves basis with energy cutoff 450 eV was used. The convergence for energy in the process of structural optimization by descent methods between two successive steps was $10^{-8}$ eV.

## 3. Static bi-structure in the L-J model

Here we give some comments on the static bi-structure of the Lennard-Jones chain which was found in [16]. This structure appears in the *equidistant* L-J chain, which corresponds to the cumulene state of carbyne, above a certain critical value of the strain. The symmetry of the bi-structure is determined by inversion in its center, and the chain in this state represents two *subchains* with equal short bonds (a), which are separated by a long bond (b) [see Fig. 1].

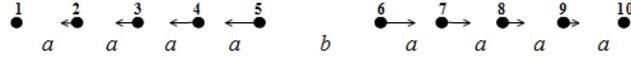

**Fig. 1.** Bi-structure of the chain with $N = 10$ particles. Atomic displacements from the initial equidistant positions are shown by arrows (*a* – short bonds, *b* – long bond).

Figure 1 shows a chain of $N = 10$ atoms with fixed ends possessing $N - 2 = 8$ degrees of freedom, which corresponds to atomic displacements along the chain. In the state of bi-structure, the chain has $N - 2 = 8$ short bond lengths (interatomic distances), which are located near the edges of the chain, and one long bond in its center. The bi-structure is symmetric about the center of the chain, which coincides with the center of a long bond.

All the short bonds are the same and shorter by some quantity $q$ ($q > 0$) than the bond lengths in the strained *equidistant* state of the chain, i.e. they are equal to $(a - q)$, where $a = a_0 * (1 + \eta)$. Here $a_0$ is the interatomic distance in the unstrained chain, and $\eta$ is the value of the strain in percentages. Thus, all atoms of the chain are displaced to its ends, and then the long bond is equal to $[a + q * (N - 2)]$. When $q = 0$, we obviously obtain a completely equidistant structure, in which all the bond lengths are equal to a.

One can say that the bi-structure consists of two identical *subchains*, which are separated from each other by one long bond. Since in the L-J model all short bonds are equal to each other, we call such bi-structure *homogeneous*.

In [16], we have considered how to construct the ideal bi-structure with the aid of displacements of all the atoms from their positions in the equidistant structure, and have shown that the set of such displacements should form an *exact* arithmetic progression with the difference $q$. Now, we discuss why the ideal bi-structure can exist in the L-J model.

Since this bi-structure is an equilibrium state of the chain, the forces acting on each of its atoms should be zero. In the L-J model, we take into account only interactions between the nearest neighbors and, therefore, the forces acting on each atom from its left and right neighbors must be equal. This condition is automatically satisfied for atoms that are at the junction of two short bonds because they are equal. Thus, we must demand the fulfillment of the above condition only for atoms that are located at the junction of the short and long bonds (atoms #5 and #6 in Fig. 1). Since the distribution of bonds in the bi-structure is symmetric with respect to the center of the chain, it is sufficient to require equilibrium of only one atom (atom #5 in Fig. 1):

$$f(a - q) = f(a + (N - 2) * q), \quad (1)$$

where $f(r)$ is the Lennard-Jones force acting between two particles at a distance $r$ from each other.

Equation (1), which we call hereafter the "force equation", is a high degree algebraic equation with many different roots $q$. We have shown in [16] how above some strain ("critical strain") a certain bifurcation occurs in the solution of this equation, which leads to the appearance of two real roots. One of these roots gives us a nonzero value of $q$, generating the stable bi-structure. This bifurcation is *hard* (similarly to a first order phase transition), i.e. the root $q$ appears abruptly with nonzero value.

In the L-J model, the critical strain ($\eta_c$) can be calculated for different $N$ with a high degree of accuracy, for example, $\eta_c$ for $N = 10$ is equal to 5.7704%.

Let us explain the physical nature of the appearance of the bi-structure as a result of solving the force equation (1). The right part of the force equation (1) is associated with the tail of the Lennard-Jones potential, while its left part is associated with a small vicinity of the minimum of this potential. Indeed, the right hand side of the force equation is *small* because its argument is a long bond, which has a sufficiently large value and, therefore, it relates to the tail of the potential. Then the left part of the force equation, whose argument is a short bond, also must be small. However, the Lennard-Jones force can be small, except the tail of the potential, only in a small vicinity of its minimum (at the point of the potential minimum it is zero). Since any realistic interatomic potential has a minimum and a tail, where it tends to zero, it is clear that the above arguments are true for it as well. In other words, the bi-structures can exist in wide classes of interatomic interactions.

## 4. Static bi-structure in the DFT-model

As was already mentioned in Introduction, the authors of the papers [4, 5] have investigated the static structures of strained finite carbon chains. They paid particular attention to the *inhomogeneity* of these structures and to dependence of this property on the length of the chains and on the parity of the number of their atoms. In some cases, quasi-polyyne structures were observed (at least in the middle part of the chain), and in some cases quasi-cumulene structures (more precisely, it is a polyyne structure with a very small difference among the bond lengths). We have found essentially different equilibrium structures, which appear above a certain strain of the carbyne, namely, *inhomogeneous bi-structures*. The fact that these structures were not found in the above papers can be explained by using the cumulene approximation (or approximation close to it) in minimizing the total energy of the system. In such a way, one misses some other energy minima, which can be found only from fundamentally different initial approximations. First, we consider the form of these bi-structures, and then discuss the procedure to obtain them.

Let us consider the bi-structure in the carbon chain of $N = 16$ atoms. The corresponding bifurcation, appears in a narrow interval of chain strain $\eta = [6.10\%; 6.11\%]$. Table 1 shows the distribution of bond lengths in the chain *before* and *after* this bifurcation, while Fig. 2 shows a graphical representation of these bond lengths (BL).

**Table 1.** Bond lengths (in angstrom) before and after bifurcation for $N = 16$ particle chain

| Bonds | $\eta = 6.10\%$ (before bifurcation) | $\eta = 6.11\%$ (after bifurcation) |
|---|---|---|
| 1-2 | 1.421 | 1.311 |
| 2-3 | 1.403 | 1.309 |
| 3-4 | 1.361 | 1.279 |
| 4-5 | 1.381 | 1.305 |
| 5-6 | 1.363 | 1.272 |
| 6-7 | 1.380 | 1.321 |
| 7-8 | 1.364 | 1.276 |
| 8-9 | 1.379 | 2.582 |
| $E/N$, eV | -7.60 | -7.40 |

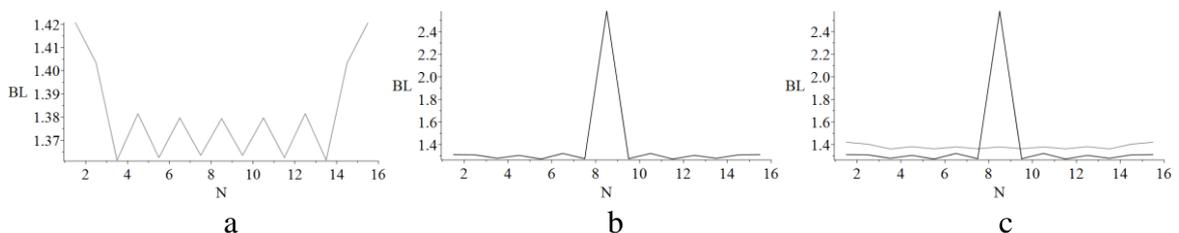

**Fig. 2.** Bond lengths before (a) and after (b) bifurcation for $N = 16$ particle chain. Bond lengths before and after bifurcation depicted in a single picture (c).

The structures that take place *before* the bifurcation are completely similar to those discussed in [4, 5]. It can be seen from Fig. 2 that in the middle of the chain there is an alternation of bond lengths, typical for the polyyne, and that the first bonds near the fixed ends of the chain are significantly longer than other bonds. We consider the structures with *different* short bonds as *inhomogeneous*, in contrast to the ideal equidistant structures, which take place in the L-J model.

As was already discussed, a new minimum of the total energy, corresponding to the bi-structure, appears when we pass through the critical value of the strain. We call it minimum 2, in contrast to the minimum 1 whose vicinity is associated with quasi-polyyne structures. It should be emphasized that both minima coexist in the system at the same time, and the choice of one of them by DFT descent method depends only on the appropriate *initial approximation*. The distance between the above minima increases rapidly with increasing of the strain.

The long bond, arising in the center of the chain with $N = 16$ atoms, is approximately twice as large as the short bonds in the bi-structure (this length depends on $N$).

Note that short bonds in the bi-structure become smaller than those taking place in the structure before bifurcation (see Fig. 2c). This can be explained by the appearance of the long bond in the middle of the chain that weakens the attraction of all the atoms to the center of the chain compared to their attraction to its ends.

In Table 2, one can see dependence of some parameters of the considered bi-structures on the number $N$ of the carbon atoms in the chains.

**Table 2.** Dependence of bi-structure's parameters on the number $N$ of particles in chains

| N | Critical strain $\eta_c$ | Long bond | Short bonds | |
|---|---|---|---|---|
| | | | Min | Max |
| 8 | 14.68 | 2.443 | 1.297 | 1.364 |
| 10 | 7.06 | 2.752 | 1.272 | 1.287 |
| 12 | 8.80 | 2.517 | 1.281 | 1.329 |
| 14 | 4.77 | 2.419 | 1.262 | 1.290 |
| 16 | 6.11 | 2.582 | 1.272 | 1.321 |
| 18 | 3.88 | 2.644 | 1.261 | 1.281 |
| 20 | 4.56 | 2.666 | 1.268 | 1.313 |
| 22 | 2.65 | 2.532 | 1.260 | 1.297 |
| 24 | 3.57 | 2.745 | 1.266 | 1.298 |
| 26 | 2.07 | 2.499 | 1.263 | 1.307 |

It is interesting to consider the energy profile when we move along the straight line in the multidimensional space connecting the two above discussed minima. Fig. 3 shows how this profile changes with increasing of the strain above its critical value, at which the minimum 2 appears. Let the coordinates of minimum 1 and minimum 2 are $(x_1, x_2, ..., x_n)$ and $(y_1, y_2, ..., y_N)$, respectively, while the current point along the straight line in $N$-dimensional space has coordinates $(z_1, z_2, ..., z_N)$. If this point lies on the line passing through our two minima, its coordinates must satisfy the equation $z_i = x_i - k * (y_i - x_i)$, $i = 1 .. N$, where $k$ is the current parameter. If $k = 0$, then $z_i = x_i$ for all $i$, i.e we get to the minimum 1. If $k = 1$, then $z_i = y_i$, i.e. we get to the minimum 2.

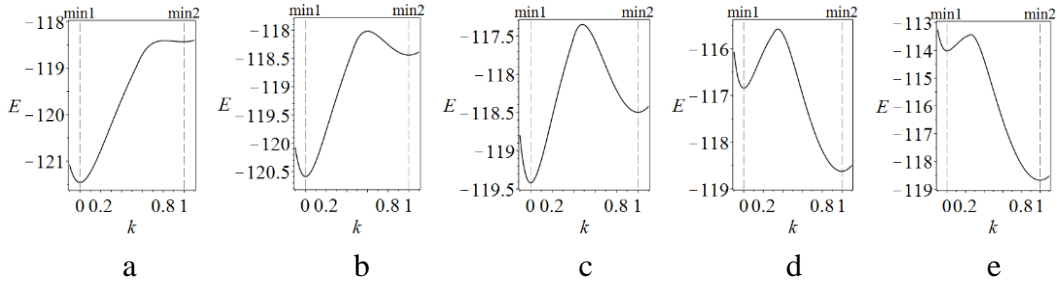

|  a | b | c | d | e |

**Fig. 3.** Energy profiles for the carbon chains with $N = 16$ atoms between two minima (see explanations in the text) for several strains: (a) 6.2%, (b) 7%, (c) 8%, (d) 10%, (e) 12%.

We show these energy profiles in Fig. 3, which demonstrates the appearance of the second minimum (min 2) for the chain with $N = 16$ atoms and a gradual increase of its depth with increasing strain. At $\eta = 12\%$, the second minimum is much lower than the traditionally considered first minimum (min1), i.e. at such a strain the bi-structure is energetically more preferable in comparison with the quasi-polyyne structure.

Let us comment on the initial approximation, which leads the descent method to the minimum 2. This approximation is prompted by our study of the L-J model.

In DFT modeling, a certain variant of descent methods (usually the conjugate gradient method) is used to find the equilibrium configuration of the considered system. But in all descent methods, it is necessary to choose an adequate initial configuration of the system in order to get to the desired minimum, if its total energy has several minima. Since we are looking for the second minimum (min2), which corresponds to the bi-structure, it is necessary to choose the initial configuration of the system close to this bi-structure. Therefore, we take two chains consisting of $N/2$ atoms with an exact *cumulene* structure, which are at a sufficiently large distance from each other. In our case, 3 Å can be chosen as this distance, because it exceeds the value of the long bond of the expected bi-structure.

With this choice of the initial configuration, a weak attraction between these chains appears and they become closer to each other, while their exact cumulene structures transform into quasi-polyyne structure typical to the finite carbon chains. Eventually, the descent method leads us to such bi-structure, as specified in the Table 1 and Fig. 2. Remember that this second minimum occurs abruptly when we go through the critical value of the strain.

## 5. Some properties of bi-structures of carbon chains

One can see from Table 2 that there is a tendency of decreasing of the critical strain $\eta_c$ with an increase in the number $N$ of the carbon atoms. However, the function $\eta_c(N)$, is not monotonic and seems to be strange. The reason of this phenomenon lies in the *parity effect*, but not in that reported in [4, 5]. Indeed, these papers discussed the parity effect corresponding to the whole chain of $N$ atoms, while our parity effect associated with the *subchains* of the chain consisting of $N/2$ atoms each. Let us consider this point in more details.

It follows from Table 2 that separately for even $N/2$ and for odd $N/2$ the functions $\eta_c(N)$ are monotonic and can be approximated by simple polynomials. We will discuss this problem, as well as the scaling of these functions for large $N$ elsewhere.

Here it is appropriate to comment on comparison of the critical strains in L-J and DFT models. The difficulty of such a comparison arises because in the L-J model we deal with homogeneous bi-structures, while in the DFT model inhomogeneous ones take place. We must choose the length $L_0$ of the chain without strain relative to which we then determine the percent of the strain. If the length of the chain under the strain $\eta$ is equal to $L_1$, then $\eta = (L_1 - L_0)/L_0$. This definition of the strain coincides with that used by us earlier for the homogeneous structures [16]: $\eta = (a - a_0)/a_0$, where $a_0$ and $a$ are the size of the primitive cell in the absence and in the presence of the strain, respectively.

In [4, 6], the change of conductivity of the carbon chains as a consequence of the Peierls phase transition is discussed. A more considerable change in the electrical properties of carbon chains should take place in the vicinity of bifurcation, after which the bi-structure appears. This is obvious, if we consider the tight-binding approach of the crystals theory. In this approach, the probability of an electron jumping from one atom to another is determined by the overlapping degree of wave functions of neighboring atoms. It is clear that this overlapping essentially depends on the distance between these atoms, which increases abruptly at the bi-structure center.

Calculations of the electron density between atoms in the DFT model shows that it decrease by several orders of magnitude in the region of a long bond. Practically, the electron density simply does not exist there. This means that there is no overlapping of the atomic orbitals in the region of the long bond. In other words, atoms in this region do not interact due to the overlapping of their shells. However, their interaction is provided by the van der Waals (vdW) forces, which decrease with the distance between atoms as $r^{-6}$.

## 6. Conclusion

The L-J model for explanation the static and dynamical properties of the *infinite* strained carbon chains has been proposed in our paper [11]. With its help, it was possible to explain the phenomenon of the pi-mode softening in some range of its amplitudes, which we found in the simulation based on the density functional theory. The condensation of this mode leads to the Peierls phase transition, which radically changes the electronic properties of carbyne [6]. With the aid of the L-J model, the possibility of existence of two new forms of carbyne, besides cumulene and polyyne, was predicted in [11].

Within the framework of the same L-J model for *finite* strained carbyne chains, we found homogeneous static bi-structures and discrete breathers of a new type that can exist in the vicinity of these structures [16].

However, L-J modeling belongs to the molecular dynamics approach with all its approximations and shortcomings. In contrast, the DFT modeling is a much more adequate method for studying the various properties of crystal structures, since it is based on the quantum-mechanical approach and allows one to take into account the deformation of atomic electron shells.

In the present paper, we study with the aid of DFT-modeling bi-structures in the finite strained carbon chains. In contrast to the homogeneous bi-structures arising in the L-J model, the similar objects in DFT models of carbon chains represent *inhomogeneous* bi-structures.

Thus, the predictions obtained on the basis of molecular dynamics methods (L-J model, in our case) can be extremely useful for studying material properties in the framework of a more realistic approach based on the density functional theory.

In conclusion, we note one curious fact, connecting with the appearance of bi-structures in the finite carbon chains. The bifurcation point in the multidimensional coordinate space is a *saddle* stationary point of the potential energy in the Lennard-Jones model, while it is a clearly expressed energy *minimum* in the framework of the density functional theory. The same situation takes place for the case of the Peierls phase transition in the infinite carbon chain, which we discussed in detail in the paper [19], despite different nature of the corresponding bifurcations (soft bifurcation in this case against hard bifurcation in the case of the present paper). Some general arguments explaining this fact can be found in [19].

**Acknowledgments**

*The authors acknowledge support by the Ministry of Science and Higher Education of the Russian Federation (state assignment grant No. 3.5710.2017/8.9) and they are sincerely grateful to N. V. Ter-Oganessian for useful discussions.*